\documentclass[%
reprint,
superscriptaddress,
showpacs,preprintnumbers,
 amsmath,amssymb,
aps,
prl,
floatfix%,
]{revtex4-1}
\usepackage{amsmath}
\usepackage{graphicx}
\DeclareMathSizes{12}{17.28}{9}{7}

\usepackage{float}
\usepackage{dcolumn}% Align table columns on decimal point
\usepackage{bm}% bold math
\usepackage{bbold}
\usepackage[multidot]{grffile}
\usepackage[colorlinks=true, breaklinks=true]{hyperref}% add hypertext capabilities
\usepackage[mathlines]{lineno}% Enable numbering of text and display math
\usepackage{comment}
\usepackage{xcolor}

\usepackage{xspace}

\usepackage[utf8]{inputenc}
\usepackage{tikz}
\usetikzlibrary{shapes.geometric,fit,arrows,babel}
\usepackage[utf8]{inputenc} % allow utf-8 input
\usepackage[T1]{fontenc}    % use 8-bit T1 fonts
\usepackage{hyperref}
\hypersetup{colorlinks,
citecolor=blue
}
\usepackage{url}            % simple URL typesetting
\usepackage{booktabs}       % professional-quality tables
\usepackage{amsfonts}       % blackboard math symbols
\usepackage{nicefrac}       % compact symbols for 1/2, etc.
\usepackage{microtype}      % microtypography
\usepackage{lipsum}
\usepackage{amsmath,amssymb,amsthm}
\usepackage[normalem]{ulem}
\usepackage[colorinlistoftodos,prependcaption,textsize=tiny]{todonotes}
\usepackage{graphicx}
\usepackage{wrapfig}

\definecolor{Agreen}{rgb}{0.1, 0.6, 0.1} % define new green
\usepackage{ulem}

\begin{document}

%\preprint{APS/123-QED}

\title{Self-induced-stochastic-resonance breathing chimeras}

\author{Jinjie Zhu}
\email{jinjiezhu@nuaa.edu.cn}
 \affiliation{State Key Laboratory of Mechanics and Control of Mechanical Structures, College of Aerospace Engineering, Nanjing University of Aeronautics and Astronautics, Nanjing 210016, China}
 
\author{Marius E. Yamakou}
\email{marius.yamakou@fau.de}
\affiliation{Department of Data Science, Friedrich-Alexander-Universit\"{a}t Erlangen-N\"{u}rnberg, Cauerstr. 11, 91058 Erlangen, Germany}
\affiliation{Max-Planck-Institut f\"ur Mathematik in den Naturwissenschaften, Inselstr. 22, 04103 Leipzig, Germany}

\date{\today}

\begin{abstract}
The study in [Phys. Rev. Lett. 117, 014102 (2016)] discovered a novel type of chimera state known as coherence-resonance chimera (CRC), which combines the effects of coherence resonance (CR) and the spatial property of classical chimeras. In this Letter, we present yet another novel form of chimera, which we refer to as self-induced-stochastic-resonance breathing chimera (SISR-BC), which differs fundamentally from the CRC in that it combines the mechanism and effects of self-induced stochastic resonance (SISR, previously shown in [Phys. Rev. E 72, 031105 (2005)] to be intrinsically different from CR), the symmetry breaking in the rotational coupling between the slow and fast subsystems of the coupled oscillators, and the property of breathing chimera --- a form of chimera state characterized by non-stationary periodic dynamics of coherent-incoherent patterns with a periodically oscillating global order parameter. Unlike  other types of chimeras, including CRC, SISR-BC demonstrates remarkable resilience to a relatively wide range of stochastic perturbations and persists even when the purely excitable system is significantly distant from the Hopf bifurcation threshold --- thanks to the mechanism of SISR--- and globally attract random distributions of initial conditions. Considering its potential impact on information processing in neuronal networks, SISR-BC could have special significance and applications.

\end{abstract}

%\pacs{05.45.Xt, 05.65.+b, 89.75.Fb}% PACS, the Physics and Astronomy
\maketitle

%%%%%%%%%%%%%%%%%%%%%%%%%%%%%%%%%%%%%%%%%%%%%%%%%%%%%%%%%%%%%%%%%%%%%%%%%
Noise is prevalent in complex systems and can originate from various sources with varying impacts on the overall dynamics of such systems. Interestingly, noise can have beneficial effects in some cases, such as improving weak signal detection, enhancing the time precision of oscillations for optimal information processing and transfer, or induce quasi-periodic oscillations that would not emerge without noise. This contradicts the common belief that noise is merely a nuisance in complex systems. Excitable systems, in particular, may exhibit distinct noise-induced dynamical effects when noise is introduced to the fast excitatory or slow recovery variable, depending on the deterministic parameter regime of the system. These effects can be counter-intuitive and lead to phenomena such as stochastic resonance (SR) \cite{longtin1993stochastic,wellens2003stochastic}, inverse stochastic resonance (ISR) \cite{tuckwell2009inhibition,yamakou2018weak}, coherence resonance (CR)\cite{pikovsky1997coherence,pisarchik2023coherence}, and self-induced stochastic resonance (SISR) \cite{muratov2005self,yu2021self,Zhu2021,yamakou2022levy,Zhu2022a,LeeDeVille2005,yamakou2019control}, etc.

In this Letter, the focus is on SISR, which occurs when a small random perturbation of the fast excitatory variable of an excitable system with strong time-scale separation results in the onset of quasi-periodic oscillations that would not emerge without noise. SISR combines a stochastic resonance-type phenomenon with an intrinsic reset mechanism, and the system requires no periodic drive. SISR differs from other noise-induced coherent behaviors, particularly CR (see \cite{LeeDeVille2005,yamakou2019control} for details), in that it arises away from bifurcation thresholds in a parameter regime where the zero-noise dynamics do not display a limit cycle nor even its precursor.

Classical chimera states refer to intriguing spatiotemporal patterns that consist of synchronized and desynchronized behavior in spatially separated domains, which emerge in networks of identical units. These patterns were initially discovered in a network of phase oscillators with a simple symmetric nonlocal coupling scheme and have since sparked a tremendous amount of theoretical investigations. Chimera states have also been observed in various real-world systems such as power grids \cite{motter2013spontaneous}, modular neural networks \cite{hizanidis2016chimera}, and the unihemispheric sleep of birds and dolphins \cite{rattenborg2000behavioral}, among others. In the context of neuronal networks, it is particularly relevant to study chimera states under conditions of excitability, especially in the case of epileptic seizures \cite{rothkegel2014irregular}. While previous studies have reported the existence of chimera states in coupled units in the oscillatory regime, their existence in purely excitable regimes remains elusive (despite the study in \cite{Semenova2016}) even with specially prepared initial conditions and coupling strengths. In this Letter, we uncover the mechanism behind a new type of chimera in a network of oscillators in a purely excitable regime, whose occurrence is even independent of this regime's strength.

In addition to classical chimeras, different forms of chimera state have been discovered --- including multicluster chimera \cite{yao2015emergence,xie2014multicluster}, traveling chimera \cite{xie2014multicluster}, amplitude chimera \cite{zakharova2016amplitude}, chimera death \cite{zakharova2014chimera}, coherence-resonance chimera \cite{Semenova2016}, breathing chimera \cite{bolotov2017breathing,Omelchenko2022,suda2020emergence} --- and extensively studied.
In a classical chimera state, the oscillators in the system are divided into two groups that behave differently: one group is synchronized and oscillates in phase, while the other group is desynchronized and oscillates out of phase. In a breathing chimera (BC), however, the behavior of the desynchronized group is more complex: instead of maintaining a fixed out-of-phase relationship with the synchronized group, some oscillators in the desynchronized group alternate between being in-phase and out-of-phase with the synchronized group. This gives the pattern a ``breathing'' quality as the out-of-phase group expands and contracts over time, forcing the coupled oscillator's global order parameter to oscillate periodically --- the prominent characteristic of the BC state \cite{bolotov2017breathing,Omelchenko2022,suda2020emergence}.

To this day, insufficient attention has been devoted to the interplay between noise-induced resonance phenomena and chimera states. Thus, it is essential to investigate the interplay between these two phenomena and explore how a noise-induced resonance phenomenon may induce and/or benefit chimera states' emergence and stability. Only the interplay between CR and classical chimera, leading to the discovery of coherence-resonance chimera (CRC), has been investigated \cite{Semenova2016}. In CRC, CR is associated with spatially coherent and incoherent behavior rather than purely temporal coherence or regularity measured by the correlation time.

In this Letter, we narrow the gap between noise-induced resonance phenomena and chimera states by uncovering the emergent mechanism behind yet another intriguing form of chimera --- self-induced-stochastic-resonance breathing chimera (SISR-BC) that combines the emergent mechanism of SISR, a symmetry breaking in the rotational coupling between the fast and slow variables of the coupled oscillators, and breathing chimera.

%%%%%%%%%%%%%%%%%%%%%%%%%%%%%%%%%%%%%%%%%%%%%%%%%%%%%%%%%%%%%%%%%%%%%%%%%
As a paradigmatic model with well-known biological relevance, we consider a ring of $N$-coupled FitzHugh-Nagumo (FHN) neurons~\cite{fitzhugh1960thresholds,fitzhugh1961impulses,nagumo1962active}:
%\begin{widetext}
\begin{eqnarray}\label{eq:1}
\begin{split}
\left\{\begin{array}{lcl}
\displaystyle{\varepsilon \frac{dv_i}{dt}}&=&\displaystyle{v_i-\frac{v_i^3}{3}-w_i}+\frac{\sigma}{2R}\sum\limits_{j=i-R}^{i+R}\big[b_{vv}\big(v_j-v_i\big)\\[5.0mm]
&+& b_{vw}\big(w_j-w_i\big)\big]+ \xi_i(t),\\[3.0mm]
\displaystyle{\frac{dw_i}{dt}}&=& \displaystyle{v_i + a_i} +\frac{\sigma}{2R}\sum\limits_{j=i-R}^{i+R}\big[b_{wv}\big(v_j-v_i\big)\\[5.0mm]
&+& b_{ww}\big(w_j-w_i\big)\big],
\end{array}\right.
\end{split}
\end{eqnarray}
%\end{widetext}
where $v_{_{i}}\in\mathbb{R}$ and $w_{_{i}}\in\mathbb{R}$ represent the fast membrane potential and the slow recovery current variables of the elements, respectively. The index $i=1,...,N$ stands for the node $i$ in the ring network, where all indices are modulo $N$. The parameter $R$ denotes the number of nearest neighbors, and $r=R/N$ is the coupling range. The coupling strength between neurons is given by $\sigma>0$, and $0<\varepsilon\ll1$ is the timescale separation ratio between $v_i$ and $w_i$.
The parameter $a_i$ determines the excitability threshold of a neuron: if $\lvert a_i \rvert > 1$, it's excitable; if $\lvert a_i \rvert < 1$, it's oscillatory. 
This study assumes $a_i \equiv a=1.001$, except otherwise stated. Thus, all neurons are in the excitable regime. The term $\xi_i(t)$ is modeled as independent Gaussian white noise with zero mean and correlation function $\langle \xi_i(t),\xi_j(t') \rangle=2D\delta_{ij}\delta(t-t')$, where $D$ is the noise intensity.

The coupling matrix $B = [b_{vv}, b_{vw}; b_{wv}, b_{ww}]$ encodes the coupling between the slow and fast subsystem of each neuron. In the Supplemental Material (SM)~\cite{supp}, we replicated the CRC reported in Ref.\cite{Semenova2016} with the same rotational coupling. The only difference is that noise is added exclusively to the slow subsystem (a requirement for achieving CR in FHN \cite{LeeDeVille2005}) in Ref.\cite{Semenova2016} while it is added solely to the fast subsystem in our model-- a requirement for achieving SISR in FHN \cite{LeeDeVille2005}.

%%%%%%%%%%%%%%%%%%%%%%%%%%%%%%%%%%%%%%%%%%%%%%%%%%%%%%%%%%%%%%%%%%%%%%%%%
Now, we delve into the characteristics and behavior of SISR-BC. 
If, in Eq.\eqref{eq:1}, the conditions on the interplay between a weak noise intensity $D$ and a strong timescale separation ratio $\varepsilon$ required for the emergence of SISR are satisfied \cite{LeeDeVille2005,Yamakou2018} at a suitable coupling strength $\sigma$, coupling range $r$, bifurcation parameter $a$, and rotational coupling matrix $B$, then the oscillators will oscillate quasi-periodically and will become what we shall refer in the paper as SISR-oscillators.

\begin{figure}[tbp]
    \centering
    \includegraphics[width=8.5cm]{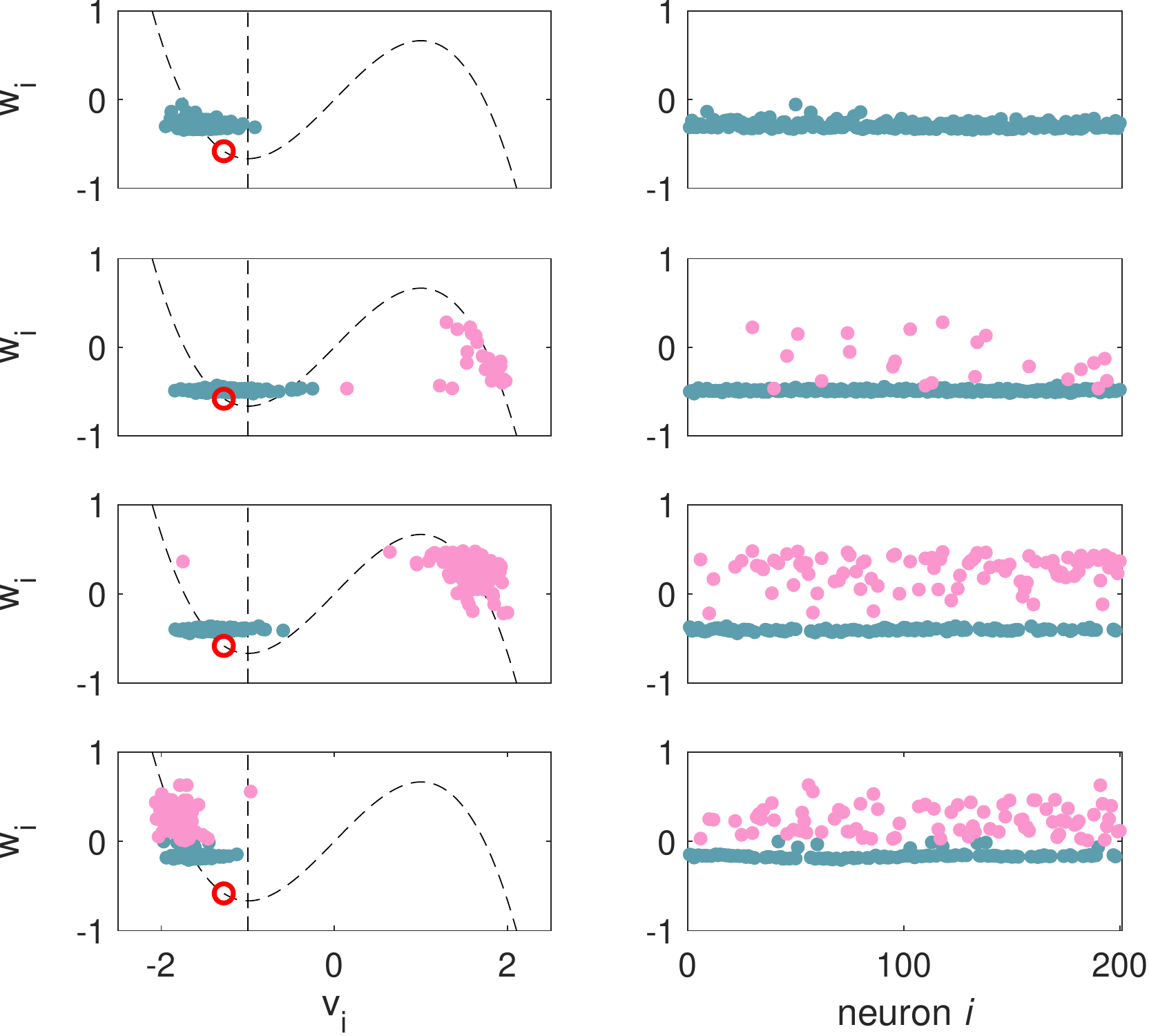}
    \caption{Time evolution of SISR-BC. Left column: snapshots of SISR-BC in phase space from top to bottom: $t=981.72, 982.24, 982.78, 983.40$ (see \cite{supp} for the full animation). Right column: Corresponding snapshots of the slow variables $w_i$. The coherent group (in blue) is restricted to $v_i<0$, $w_i\leq0$, and the incoherent group (in pink) oscillates quasi-periodically. The dashed lines are nullclines of a single FHN neuron. The red circle is the critical transition point $(v_{\rm TP},w_{\rm TP})=(-1.277,-0.583)$. Initial conditions: randomly distributed on the circle $v^2+w^2=1$. Parameters: $a=1.001$, $D=0.04$, $\varepsilon=0.01$, $\sigma=3.0$, $r=0.35$, $[b_{vv}, b_{vw}; b_{wv}, b_{ww}]=[0,0;0,1]$.}
    \label{fig:1s}
\end{figure}

By breaking the symmetry of rotational coupling such that $B = [0,0;0,1]$ and setting the parameters to, e.g., $r=0.35$, $\sigma=3.0$, $a=1.001$, $D=0.04$, and $\varepsilon=0.01$, we can observe SISR-BC. Figure \ref{fig:1s} shows four snapshots in the temporal evolution of SISR-BC (an animation of SISR-BC in phase space can be watched from SM~\cite{supp}). In the initial stage (top row of Fig.\ref{fig:1s}), after a sufficient transient time, all the oscillators (represented by blue dots) move synchronously along the left stable branch of the $v$-nullcline towards the transition point (TP) represented by the red circle in Fig.\ref{fig:1s}. The location of TP can be computed via the distance matching condition~\cite{Zhu2021, Zhu2022, Zhu2022a}. TP represents the most probable point at which a single SISR-oscillator can escape from the left to the right branch of the $v$-nullcline, and it is located to the left of the unique stable fixed point at ($v_{fp},w_{fp})=(-1.001,-0.667)$. Therefore, the escape of oscillators from the left to the right branch of the $v$-nullcline can only occur via the mechanism of SISR and not CR. We note that during CR, the escape of trajectories can only occur if the oscillators reach the minimum of the $v$-nullcine at the Hopf bifurcation at $v_{\rm min}=-1$ (see \cite{LeeDeVille2005} and SM~\cite{supp} for details).

As the population approaches TP (2nd row in Fig.\ref{fig:1s}), some (but not all) oscillators use the SISR mechanism to transit to the right branch of the $v$-nullcline at different times, making them incoherent (asynchronous) in phase space as they arrive and move up along that branch (pink dots in the 2nd row in Fig.\ref{fig:1s}).

In the 3rd row of Fig.\ref{fig:1s}, pink asynchronous SISR-oscillators start to escape from the right branch of the $v$-nullcline back to the left branch. The oscillators unable to escape remained coherent (synchronized) on the left stable branch of the $v$-nullcline, represented by the blue cluster of oscillators in the 2nd, 3rd, and 4th rows of Fig.\ref{fig:1s}. The rotational coupling matrix has a symmetry-breaking property that only favors coupling in the slow subsystem, i.e., $B=[0,0;0,1]$. Consequently, the blue cluster of coherent oscillators is pulled up (to slightly higher values of $w_i$) along the left branch of the $v$-nullcline by the pink SISR-oscillators arriving from the right branch. This upward displacement of the blue cluster of coherent oscillators can be observed by comparing its position in the 2nd and 4th rows of Fig.\ref{fig:1s}: the blue cluster is positioned further away from the red circle in the 4th row than in the 2nd row. Eventually, the pink and blue clusters merge into one cluster, and the scenario depicted in the top row of Fig.\ref{fig:1s} re-emerges, repeating the process quasi-periodically --- thanks to the SISR. It is worth noting that the coherent cluster of blue oscillators is not static in the excitable regime but instead exhibits sub-threshold oscillations as it slightly moves up and down the left stable branch of the $v$-nullcline.

Extensive numerical simulations (see SM \cite{supp}) have demonstrated that SISR-BC can only be induced by one specific form of the rotational coupling matrix, namely, $B=[0,0;0,1]$. The simulations also revealed that alternative forms of $B$ (see SM \cite{supp}) could only result in one cluster of asynchronous oscillators, two clusters in oscillation death, one cluster switching between synchrony and asynchrony, or one cluster of resting oscillators.

In the absence of a Hopf bifurcation (i.e., when $a>1$), the SISR mechanism drives the oscillators into (quasi-) periodic oscillations. This, in turn, induces the non-stationary (quasi-) periodic dynamics of coherent-incoherent patterns and a periodically oscillating global order parameter, which is the defining characteristic of breathing chimera \cite{suda2020emergence}. Thus, SISR-BC combines the mechanism and effect of SISR with the rotational coupling symmetry-breaking.

Unlike classical chimera states in CRC, which is characterized by the coexistence of spatially coherent and incoherent domains in a coupled network of oscillators ~\cite{Kuramoto2002,Abrams2008,Majhi2019,OmelChenko2018}, the coherent and incoherent groups of oscillators in SISR-BC are separated not spatially in the configuration space but in the phase space. More precisely, as shown in the 2nd column of Fig.\ref{fig:1s}, the oscillators in the coherent blue group do not necessarily synchronize with their neighbors in the configuration space (i.e., the spatial domain representing the ring network) but instead synchronize with the neighbors in the phase space (i.e., the neighbors on the left branch of the $v$-nullcline). Thus, the local order parameter defined and used to identify CRC in Ref.~\cite{Omelchenko2013,Semenova2016} (see also in SM~\cite{supp}) cannot serve as an indicator of the existence of SISR-BC.

\begin{figure}[tbp]
    \centering
    \includegraphics[width=7 cm]{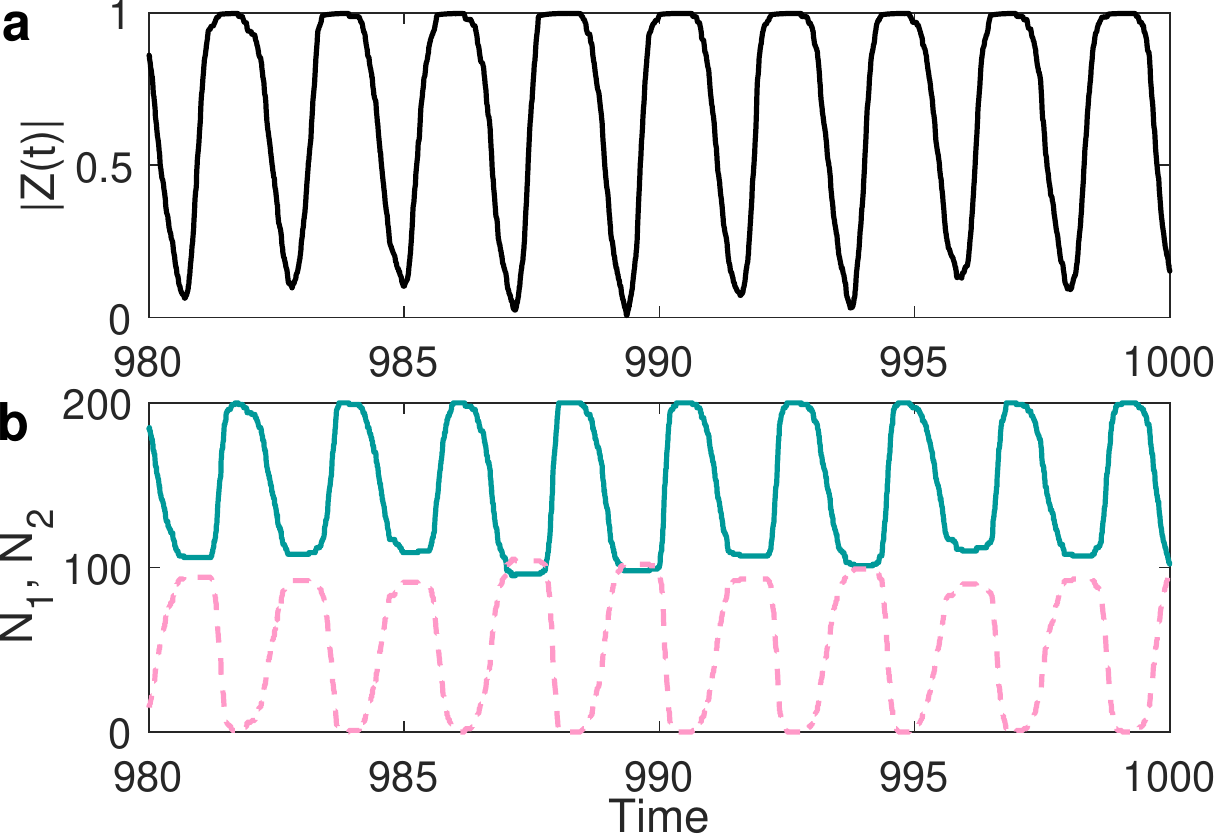}
    \caption{(a) Time series of $\lvert Z(t)\rvert$. (b) Number of oscillators in coherent (blue solid) and incoherent (pink dashed) groups. Parameter values are the same as those in Fig.\ref{fig:1s}.}
    \label{fig:2s}
\end{figure}

However, during SISR-BC, the periodic separation and aggregation of the coherent and incoherent groups (as depicted in Fig.\ref{fig:1s}) and the variation in group size of the coherent and incoherent groups (see Fig.\ref{fig:2s}(b)) force the modulus of the global order parameter now given as $Z(t)=N^{-1}\sum_{i=1}^{N}\exp(I\theta_i)$ to fluctuate with a large amplitude. Thus, the large fluctuation of $\lvert Z(t)\rvert\in[0,1]$ is an excellent indicator of the existence of SISR-BC. 

Breathing chimera states are characterized by non-stationary macroscopic dynamics of coherent-incoherent patterns that occasionally emerge in coupled oscillator networks via Hopf bifurcation \cite{suda2020emergence}. However, in the system described by Eq.\eqref{eq:1}, Hopf bifurcation is not present since the Hopf bifurcation parameter is fixed in an excitable regime ($a=1.001$). 
Nonetheless, the SISR mechanism can play the role of the Hopf bifurcation in the emergence of breathing chimera since it can induce a limit cycle behavior (quasi-periodic oscillations) in the pink group of oscillators. The main features (and hence indicators) of breathing chimera include (i) periodically varying (breathing) oscillator dynamics and (ii) oscillating modulus of the global order parameter $Z(t)$~\cite{Abrams2008,Omelchenko2022,suda2020emergence}. Thus, the non-stationarity of the coherent blue group and the incoherent pink group of SISR-oscillators (shown in Figs.\ref{fig:1s} and \ref{fig:2s}(b)), and the (quasi-) periodically oscillating modulus of the global order parameter $Z(t)$ (see Fig.\ref{fig:2s}(a)) constitute a breathing chimera state, hence the term SISR-BC. 

\begin{figure}[tbp]
    \centering
    \includegraphics[width=8.5cm]{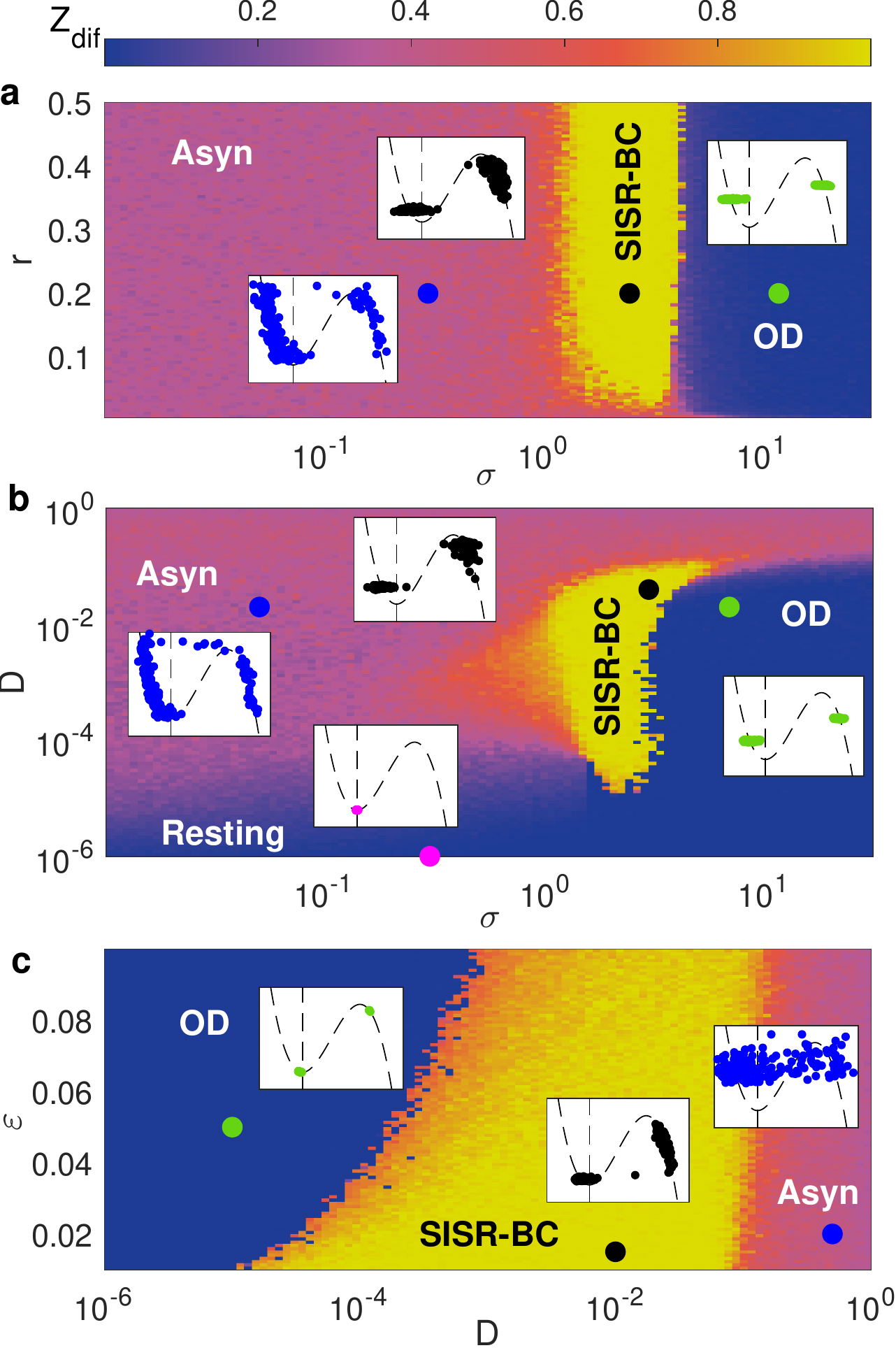}
    \caption{Dynamic regimes in the ($\sigma,r$), ($\sigma,D$), and ($D,\varepsilon$) planes. See the main text for explanations. Parameters: (a) $\varepsilon=0.01$, $D=0.04$; (b) $\varepsilon=0.01$, $r=0.2$; (c) $\sigma=2.0$, $r=0.2$, $[b_{vv}, b_{vw}; b_{wv}, b_{ww}]=[0,0;0,1].$}
    \label{fig:3s}
\end{figure}

Chimera states are well-known for their high sensitivity to different initial conditions and parametric perturbations. Small discrepancies in the system's parameters and/or initial conditions can result in the emergence of different chimera patterns or the absence of a chimera state altogether. Thus, in the rest of this Letter, we investigate the occurrence and  robustness of SISR-BC to different initial conditions and parametric perturbations. To do so, we run the numerical simulations in three relevant parameter planes, namely ($\sigma,r$), ($\sigma,D$), and ($D,\varepsilon$) shown in Fig.\ref{fig:3s}, respectively. In these parameter spaces, we performed $200$ time units of simulations and discarded the first half to avoid transients, and computed $Z_{\rm dif}=\max(|Z(t)|)-\min(|Z(t)|)$, where $\max(|Z(t)|)$ and $\min(|Z(t)|)$ are the maximum and minimum of $|Z(t)|$ in each time interval $t_i$.

In all panels of Fig.\ref{fig:3s}, the initial conditions are randomly distributed on $v^2+w^2=1$. In panel (a), we investigate the effect of varying the coupling range $r$ and coupling strength $\sigma$ on the occurrence of SISR-BC while keeping the other parameters ($D$, $\varepsilon$, and $a$) constant. The parameter range in panel (a) reveals three distinct collective behaviors: asynchrony (Asyn), SISR-BC, and oscillation death (OD). The yellow region indicates parameter values that induce SISR-BC. The snapshots of the phase portrait for Asyn, SISR-BC, and OD correspond to the system's behaviors at parameter values associated with the colored dots (purple, black, and green, respectively) indicated. Remarkably, SISR-BC remains resilient to changes in $r$ and can occur and persist in both non-locally ($0.05\lesssim r<0.5$) and globally coupled ($r=0.5$) networks, provided $\sigma$ is optimal, i.e., neither too weak nor too strong.

In Fig.\ref{fig:3s}(b), we explore the $(D,\sigma)$ plane by fixing $r=0.2$ chosen from Fig.\ref{fig:3s}(a) for inducing SISR-BC. Up to four collective behaviors can now occur: Asyn, resting, SISR-BC, and OD. The resonant aspect of SISR-BC is evident from intermediate values of the noise intensity $D$ and the coupling strength $\sigma$ necessary to induce SISR-BC. It's worth pointing out the existence of multistability, as the system can exhibit either OD or resting states for small enough $D$ and large $\sigma$, depending on the initial conditions. However, since our focus is SISR-BC, we do not extend our discussion to these other collective behaviors.

Figure~\ref{fig:3s}(c) explores the $(\varepsilon,D)$ parameter space with suitable (for SISR-BC) values of $r=0.2$ and $\sigma=2.0$ chosen from Figs.\ref{fig:3s}(a) and (b). The simulation reveals three collective behaviors: Asyn, SISR-BC, and OD. It is important to note that the occurrence and degree of SISR principally depend on an interplay between $D$ and $\varepsilon$. For small values of $D$, as $\varepsilon$ decrease, the degree of SISR becomes stronger, and trajectories are more likely to escape from the left stable branch of the $v$-nullcline above the unique stable fixed point. While we won't delve into the details of SISR in this discussion (see \cite{Yamakou2018, Zhu2021, Zhu2022a,supp}), our primary focus is on SISR-BC. In Fig.\ref{fig:3s}(c), we observe that as the degree of SISR becomes stronger(i.e., as $\varepsilon$ decreases), the yellow region representing SISR-BC becomes broader and brighter, indicating a more pronounced degree of SISR-BC. Moreover, when the noise becomes too weak at a small but fixed value of $\varepsilon$, the system falls into OD. This confirms that the mechanism of SISR is one of the baseline phenomena for this type of noise-induced chimera state.

The results presented in Fig.\ref{fig:3s} were obtained using random initial conditions on the unit circle, where $v^2+w^2=1$. Extensive simulations (see SM for initial conditions uniformly and non-uniformly distributed on the unit disk $v^2+w^2\leq1$~\cite{supp}) indicate that SISR-BC is remarkably resilient to changes in initial conditions, as the initial conditions can be unimportant to chimeras in the presence of noise as shown in \cite{Loos2016,Zhang2020}.

\begin{figure}[tbp]
    \centering
    \includegraphics[width=8.5cm]{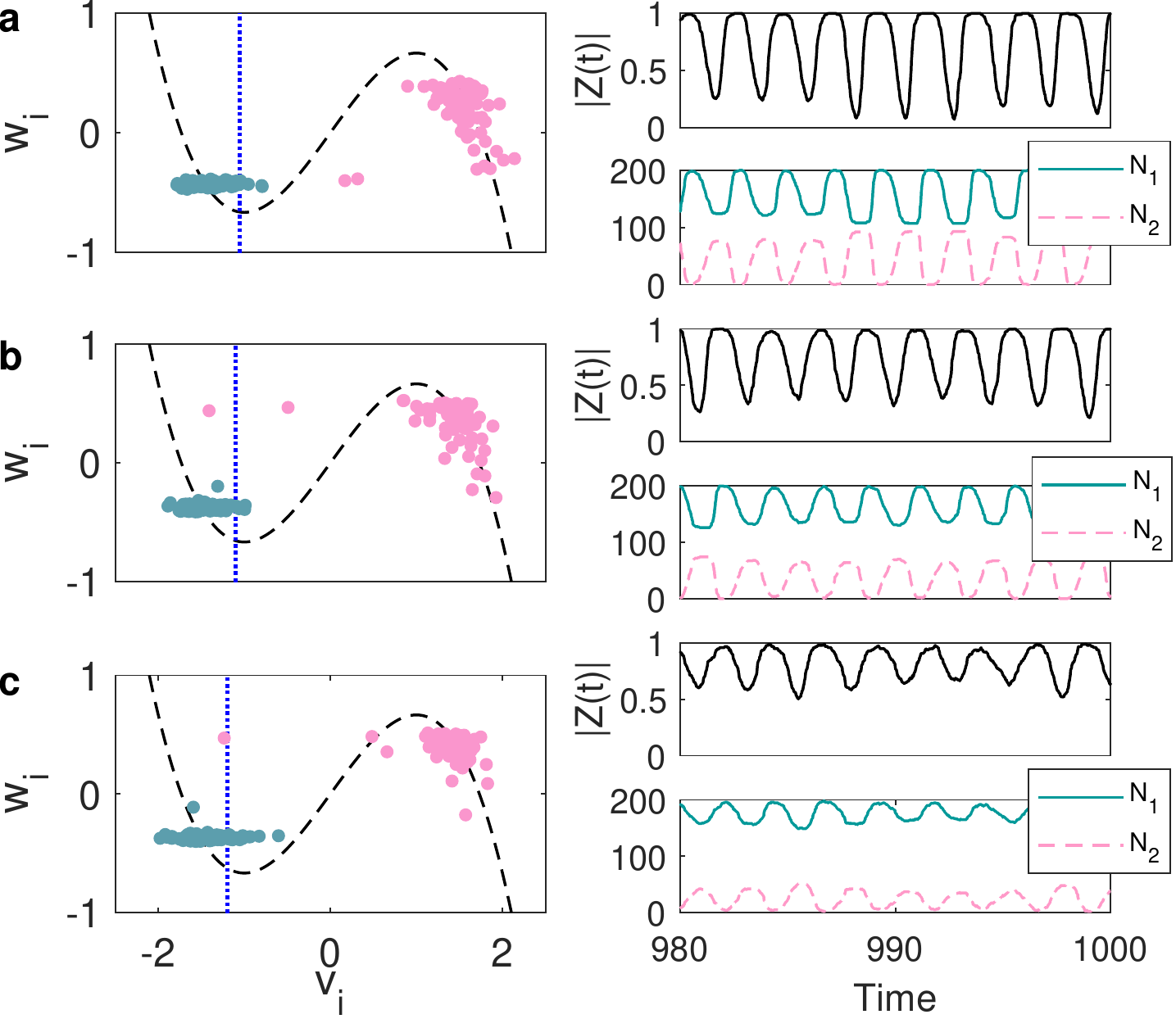}
    \caption{Snapshots of SISR-BC in phase space alongside the corresponding time series of $\lvert Z(t)\rvert$, the number of oscillators in the coherent blue group $N_1$ and incoherent pink group $N_2$. SISR-BC persists even in strong excitable regimes. (a) $a=1.05$, (b) $a=1.1$, and (c) $a=1.2$. Other parameters: $\varepsilon=0.01$, $\sigma=3.0$, $r=0.35$, $D=0.04$, $[b_{vv}, b_{vw}; b_{wv}, b_{ww}]=[0,0;0,1].$}
    \label{fig:4s}
\end{figure}

In contrast to CRC, which is observed in a network of FHN oscillators only when the Hopf bifurcation parameter $a$ is very close to the bifurcation threshold at $a=1$ (i.e., in the range $0.995 \leq a \leq 1.004$) \cite{Semenova2016}, SISR-BC can occur even when the FHN oscillators are far away from the Hopf bifurcation threshold. Figure~\ref{fig:4s} shows SISR-BC for three different values of $a$ in the excitable regime: $a=1.05, 1.1$, and $1.2$. As the value of $a$ increases, the amplitude of $|Z(t)|$ decreases, indicating a reduction in the number of oscillators in the incoherent group (in pink), leading to an OD or resting state if $a$ becomes too large.

%%%%%%%%%%%%%%%%%%%%%%%%%%%%%%%%%%%%%%%%%%%%%%%%%%%%%%%%%%%%%%%%%%%%%%%%%
In summary, the existence of self-induced-stochastic-resonance breathing chimera (SISR-BC) patterns is determined by the combined effects of the mechanism of SISR (which relies on the interplay between the noise intensity and timescale separation ratio) and the symmetry-breaking of rotational coupling between the slow and fast subsystems of the oscillators. The SISR attribute of SISR-BC induces periodic spiking (supra-threshold oscillations) of some (but not all) oscillators, although the deterministic network is entirely in an excitable regime. Meanwhile, the rotational coupling symmetry-breaking attribute of SISR-BC (i) splits the neural population into two groups --- the non-spiking group (not affected by the SISR mechanism, which stays coherent (synchronized) with sub-threshold oscillations in the excitable regime) and the spiking group (affected by the SISR mechanism with supra-threshold oscillations in the oscillatory regime) and (ii) affects the spiking time of the oscillators in the spiking group. These combined effects result in the appearance and co-existence of coherent and incoherent oscillators in the phase space (instead of the configuration space as in the traditional chimeras). The periodicity of spiking activity induced by the SISR mechanism leads to non-stationary periodic dynamics of coherent-incoherent patterns with a periodically oscillating global order parameter that characterizes a breathing chimera state. 
Furthermore, unlike some forms of chimera, we have demonstrated the robustness of SISR-BC to initial conditions. Also, we have shown that, unlike CRC, SISR-BC persists in parameter regimes far away from the Hopf bifurcation threshold.

Since we used a paradigmatic stochastic model for neural excitability, we anticipate even broader applications to neuronal networks and other stochastic excitable systems in biology, chemistry, physics, and engineering. In particular, our noise-based control method presents a different approach to chimera control, in addition to the one in \cite{Semenova2016}, supplementing the existing deterministic control techniques.

%\begin{acknowledgments}
JZ acknowledges the support from the National Natural Science Foundation of China (Grant No. 12202195).
MEY acknowledges support from the Deutsche Forschungsgemeinschaft (DFG, German Research Foundation) -- Project No. 456989199.
%\end{acknowledgments}
%
%\bibliography{references}% Produces the bibliography via BibTeX.

\end{document}